\documentclass[slac_one]{revtex4}
\usepackage{graphicx}
\usepackage{fancyhdr}
\def\ie{{\it i.e.}}
\def\eg{{\it e.g.}}

\def\mpl{\ifmmode M_{pl}\else $M_{pl}$\fi}
\def\mpl{\ifmmode \overline M_{Pl}\else $\bar M_{Pl}$\fi}
\def\to{\rightarrow}

\def\atversim#1#2{\lower0.7ex\vbox{\baselineskip\zatskip\lineskip\zatskip
  \lineskiplimit 0pt\ialign{$\matth#1\hfil##\hfil$\crcr#2\crcr\sim\crcr}}}

\begin{document}
\rightline{\vbox{\halign{&#\hfil\cr
&SLAC-PUB-11113\cr
&April 2005\cr
}}}

\title{{\small{2005 International Linear Collider Workshop - Stanford,
U.S.A.}}\\ 
\vspace{12pt}
Collider Signatures of Higher Curvature Gravity} 

\author{Thomas ~G.~Rizzo}
\affiliation{SLAC, Stanford, CA 94025, USA}

\begin{abstract}
We explore the phenomenological implications at colliders for the existence 
of higher-curvature gravity as extensions to both the Randall-Sundrum(RS) and 
Arkani-Hamed, Dimopoulos and Dvali(ADD) scenarios. Such terms are expected to 
arise on rather general grounds from ultraviolet completions of General 
Relativity, \eg, from string theory. In the Randall-Sundrum model these terms 
shift the mass spectrum and couplings of the graviton tower. In the case of 
ADD they can lead to a threshold for the production of long-lived black holes. 
\end{abstract}

\maketitle

\thispagestyle{fancy}

\section{Introduction}

The Einstein-Hilbert(EH) action 
\begin{equation}
S=\int d^{4+n}x ~\sqrt {-g}~\Bigg[{M_*^{n+2}\over {2}} R-\Lambda \Bigg]\,,
\end{equation}
is the basis for General Relativity(GR) in 4d as well as the ADD{\cite {ADD}} 
and RS{\cite {RS}} models in extra dimensions. As is well known 
EH is at best an effective 
action below the scale $M_*$. As energies approaching $M_*$ are reached
additional terms may be generated in the effective action arising 
from the UV-completion of GR, \eg, string theory. If $M_*$ is not far above 
the TeV scale then these additional terms may make their presence known at 
future colliders: the LHC and ILC. From a bottom-up point of view it is not 
so clear what form such terms might take as the number of possibilities is 
vast and so we need some guidance. If we require that the new terms do not 
produce ghosts in the graviton sector, allow unitarity to be maintained and 
are `string-motivated', we arrive at a rather unique set of terms called 
Lovelock invariants{\cite {me}}.   

These Lovelock invariants come in fixed order, $m$, which we denote here 
as ${\cal L}_m$,  
that describes the number of powers of the curvature tensor, contracted in 
various ways, out of which they are constructed. Apart from normalization 
factors we can express the ${\cal L}_m$ as 
\begin{equation}
{\cal L}_m \sim \delta^{A_1B_1...A_mB_m}_{C_1D_1...C_mD_m}~R_{A_1B_1}
~^{C_1D_1}.....R_{A_mB_m}~^{C_mD_m}\,,
\end{equation}
where $\delta^{A_1B_1...A_mB_m}_{C_1D_1...C_mD_m}$ is the totally 
antisymmetric product of Kronecker deltas and $R_{AB}~^{CD}$ is the 
$D$-dimensional curvature tensor. Fortunately, as can be seen  
by this definition, the 
number of such invariants that can exist in any given dimension $D=4+n$ 
is highly 
constrained. For a space with an even number of dimensions, $D=2m$, the 
Lovelock invariant is a topological one and leads to a total derivative, \ie, 
a surface term, in the action.  All of the higher order invariants, 
$D\leq 2m-1$, vanish identically. For 
$D\geq 2m+1$, the ${\cal L}_m$ are dynamical objects that once 
added the action can alter the field equations normally 
associated with the EH term. However it can be shown that the addition of 
any or all of the ${\cal L}_m$ to the EH action still results in a theory 
with only second order equations of motion as is the case 
for ordinary Einstein gravity. Furthermore, variation of the new action 
leads to modifications of Einstein's equations by the addition of 
new terms which are second-rank symmetric tensors with vanishing covariant 
derivatives and which depend only on the metric and its first and second 
derivatives, \ie, they have the same general properties as the Einstein 
tensor itself but are higher order in the curvature. It is these benign 
properties which provide the Lovelock invariants their unique features. 

From the discussion above we see that the most general Lovelock theory in 
4-d is just EH! In 5-d, as in the RS case, all of the 
${\cal L}_{m \geq 3}$ vanish as in 4-d but ${\cal L}_2$, which is the 
Gauss-Bonnet invariant, can now  
contribute. The generalization is now quite clear: 
for $D=5,6$ only ${\cal L}_{0-2}$ can be present. For $D=7,8$ only 
${\cal L}_{0-3}$ can be present while for $D=9,10$ only ${\cal L}_{0-4}$. 
Since the ADD model assumes that the compactified space is flat, \ie, a 
toroidal compactification is assumed, the 
coefficient of ${\cal L}_{0}$ is taken to be zero in this case.
Thus for either the RS or ADD models, there are at most three new pieces 
to add to the EH action and so the generalized form of the action 
can be taken to be 
\begin{equation}
S=\int d^{4+n}x ~\sqrt {-g}~\Bigg({M_*^{n+2}\over {2}} \Bigg[R+
{\alpha\over {M_*^2}}{\cal L}_2+{\beta \over {M_*^4}}{\cal L}_3+
{\gamma \over {M_*^6}}{\cal L}_4\Bigg]-\Lambda\Bigg)\,,
\end{equation}
where $\alpha$, $\beta$ and $\gamma$ are dimensionless coefficients. 
Explicit expressions for the ${\cal L}_m$ are given elsewhere{\cite {me}}.

\section{Influence on the Randall-Sundrum Model}

How do these new terms modify the usual RS and ADD model expectations? Let us 
turn to the RS case first where only the parameter 
$\alpha$ can be non-zero. In this case a non-zero $\alpha$ will make its 
presence known by distorting the masses and couplings of the 
graviton spectrum. Recall that 
traditional RS model is based on the $S^1/Z_2$ orbifold with the metric
\begin{equation}
ds^2=e^{-2\sigma} \eta_{\mu\nu}dx^\mu dx^\nu-dy^2\,, 
\end{equation}
with $\sigma=k|y|$ defining the curvature parameter $k\sim M_*$. 
There are two branes, separated by a distance $\pi r_c$, at the orbifold 
fixed points with the SM living on one of them(the TeV brane) with only 
gravity in the bulk. The influence of a non-zero $\alpha$ on the phenomenology 
of this model is correlated with the fact that the space between the two branes 
has a large constant curvature, \ie, it is $AdS_5$. The effects of a 
non-zero $\alpha$ are found to always occur in the combination 
$\alpha k^2/M_*^2$, explicitly showing the influence of this 
curvature. Since the ratio $k^2/M_*^2$ is usually taken to be 
small(to avoid large curvature!) this damps the effects of the 
Lovelock terms to some extent. 

\begin{figure}[htbp]
\includegraphics[width=5.7cm,angle=90]{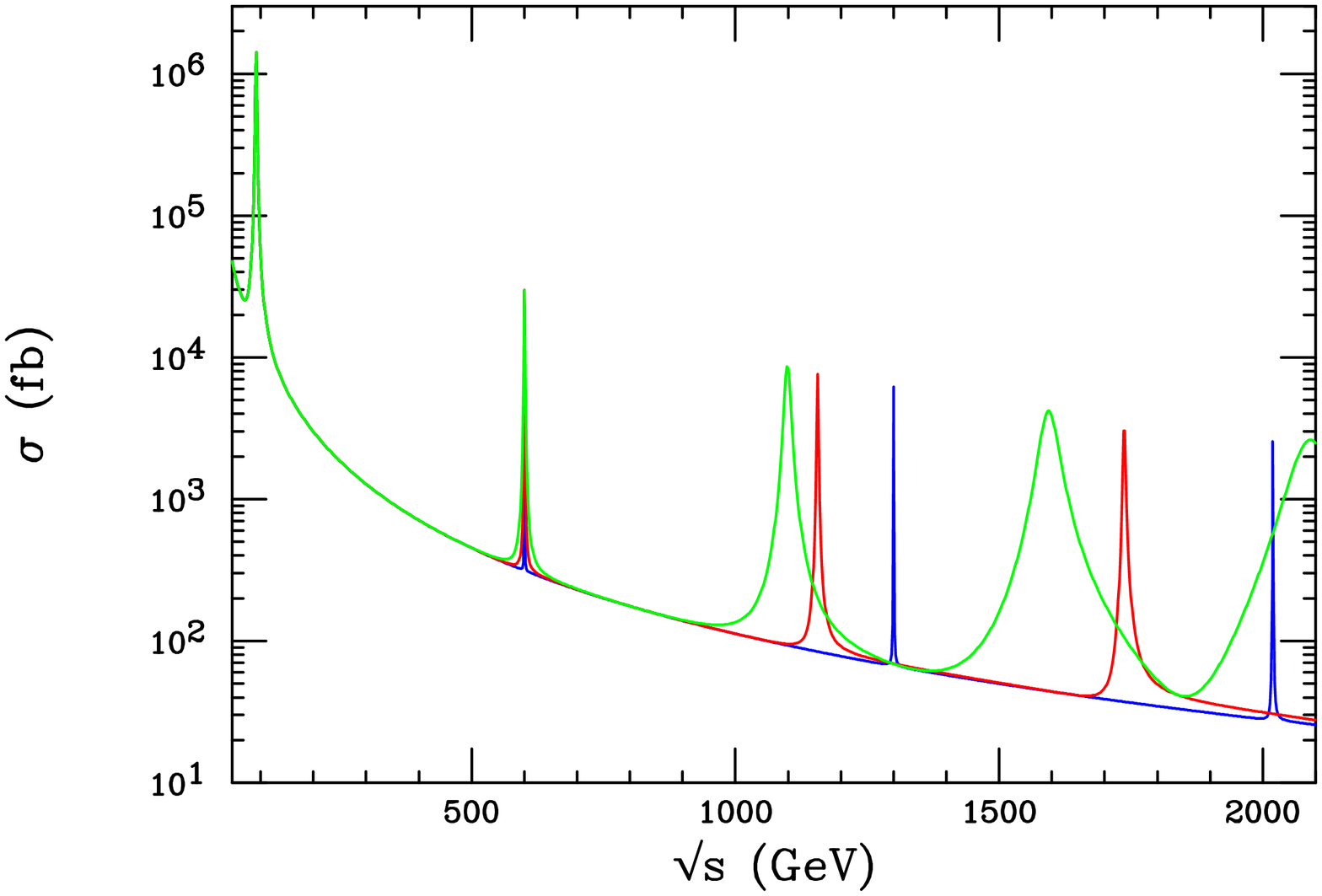}
\includegraphics[width=5.7cm,angle=90]{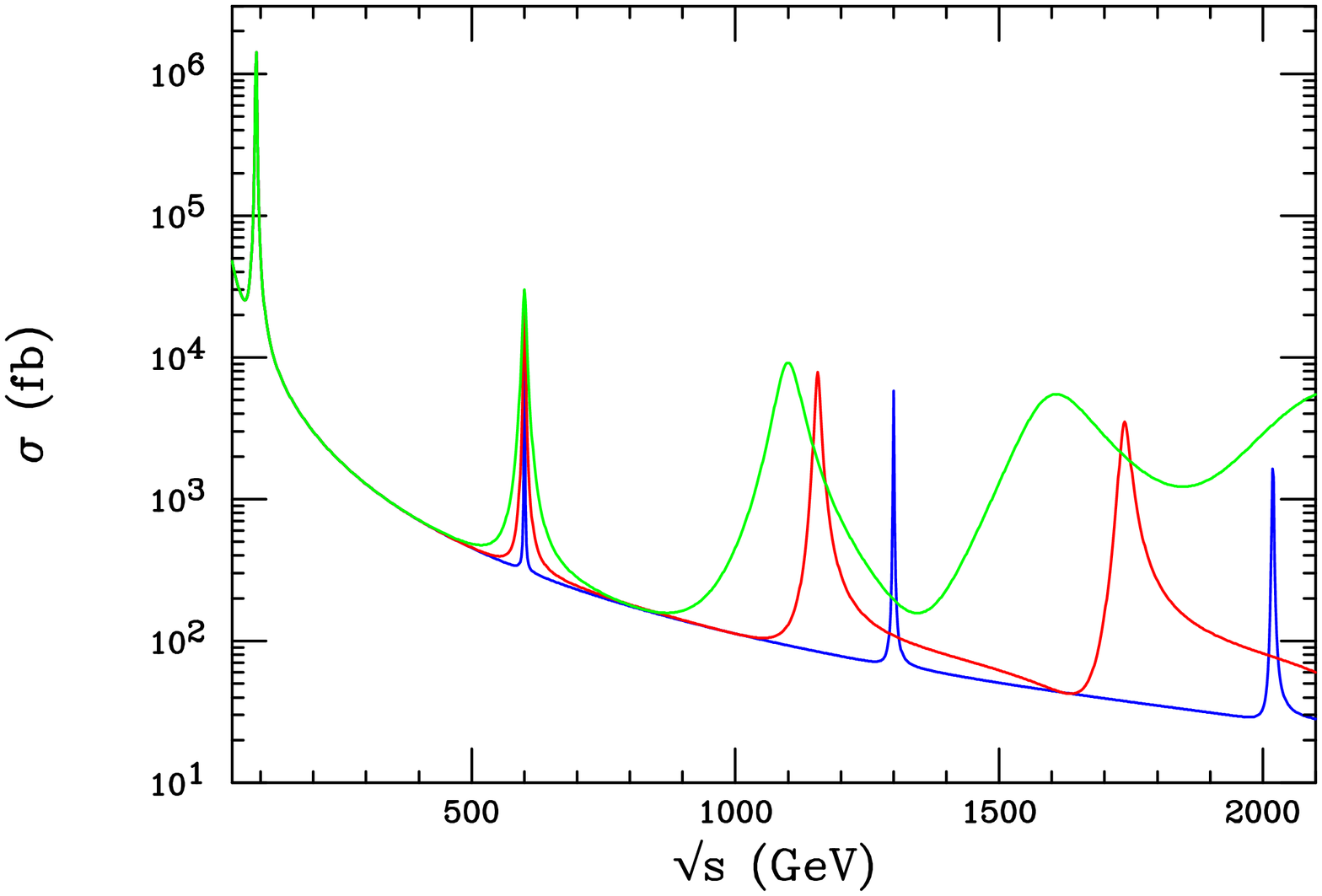}
\vspace*{0.1cm}
\caption{Cross section for $e^+e^- \to \mu^+\mu^-$ assuming $m_1=600$ GeV 
and $k/\mpl=0.05$(left) or 0.1(right). The usual RS model prediction with 
$\Omega=0$ yields the lightest spectrum(green); choosing   
$\Omega=-0.2$(red) and $-0.4$(blue) shifts the spectrum to ever larger masses.}
\label{fig1}
\end{figure}

\begin{figure}[htbp]
\includegraphics[width=5.7cm,angle=90]{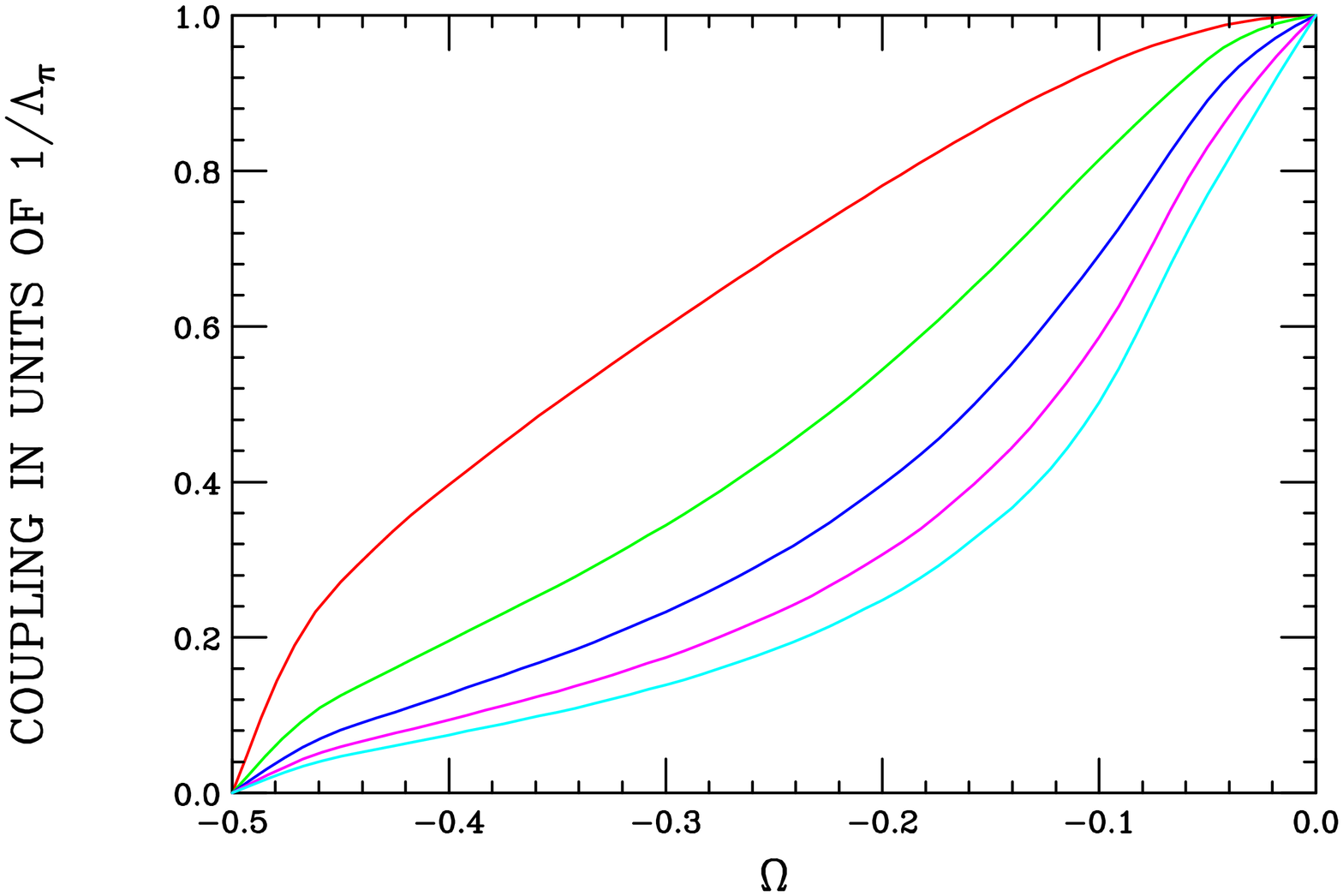}
\includegraphics[width=5.7cm,angle=90]{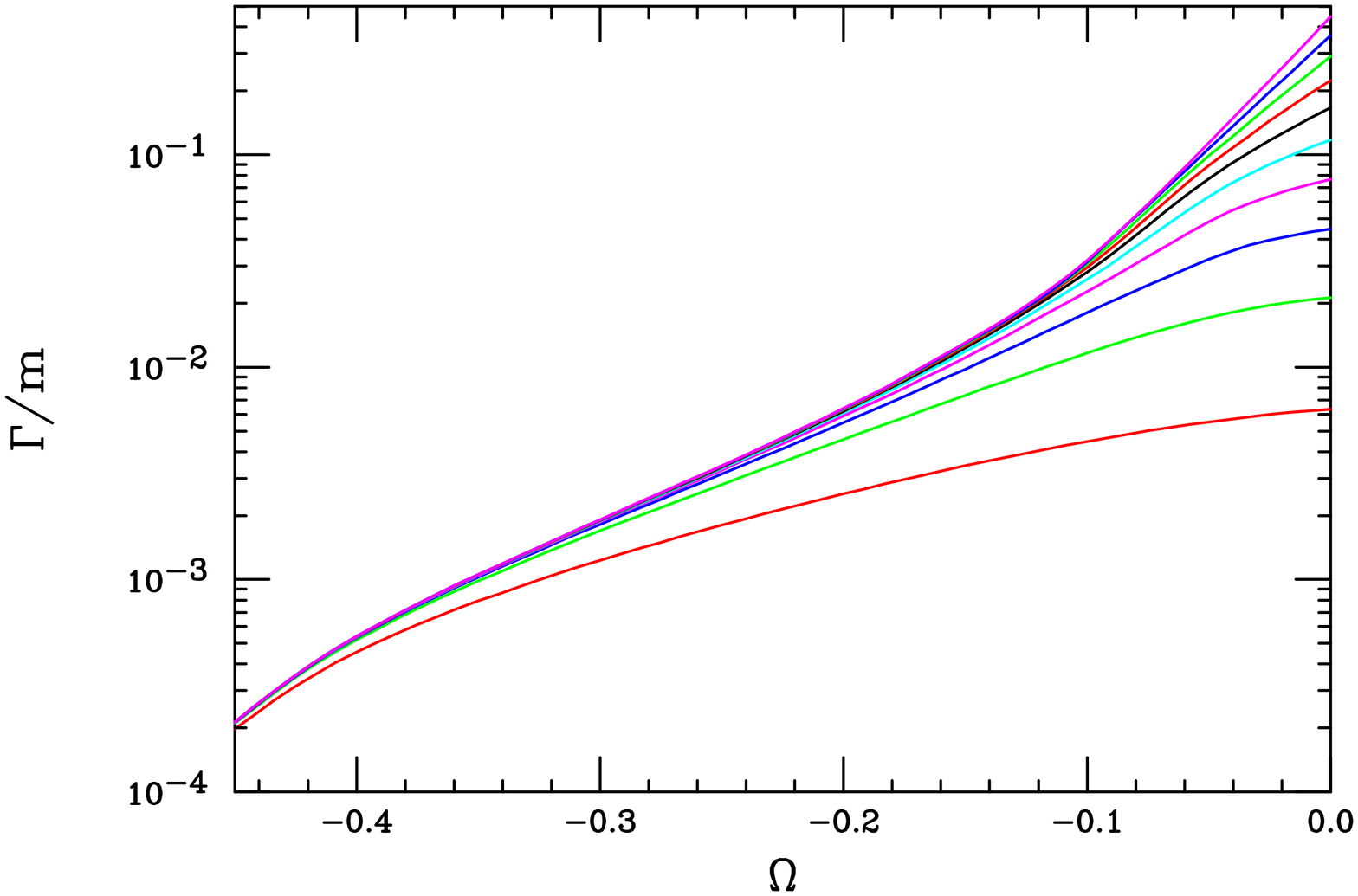}
\vspace*{0.1cm}
\caption{(Left)Coupling strengths of the first KK graviton states, from top to 
bottom, in units of 
$1/\Lambda_\pi$ as functions of the parameter $\Omega$. Note that in the RS 
limit all states have the same coupling. (Right)The ratio $\Gamma/m$ as a 
function of $\Omega$ for the first ten KK 
states. The KK number goes up as we go from the bottom to the top of the 
figure. $k/\mpl=0.05$ has been chosen for purposes of demonstration.}
\label{fig2}
\end{figure}

Defining the useful combination 
\begin{equation}
\Omega={4\alpha k^2/M_*^2\over{1-4\alpha k^2/M_*^2}}\,,
\end{equation}
it can be shown that $-1/2 \leq \Omega \leq 0$ is required to forbid tachyons 
in the Kaluza-Klein(KK) 
graviton spectrum; this forces $\alpha \leq 0$. One finds that 
$\Omega \neq 0$ causes a shift in the usual RS mass 
spectrum, \ie, the mass splitting between KK resonances 
increases, and induces a level dependence in the KK couplings 
to SM matter on the 
TeV brane by altering the boundary conditions on these two branes.
We find that the interaction of the KK graviton excitations 
with the SM fields on the TeV brane is now given by 
\begin{equation}
{\cal L}={1\over {\Lambda_\pi}} \sum_n  \Bigg[{{1+2\Omega}\over {1+2\Omega
+\Omega^2x_n^2}}\Bigg]^{1/2} h^{\mu\nu}_n T_{\mu\nu}\,,
\end{equation}
where as usual we define $\Lambda_\pi=\mpl e^{-\pi kr_c}$; the KK masses 
are given as usual by $m_n=x_nke^{-\pi kr_c}$ with the $x_n$ being the roots 
of an equation involving Bessel functions. 
The KK states are generally more massive and more narrow than in 
the standard RS case; this can be seen in the examples shown in 
Fig.~\ref{fig1}. 
In the limit $\Omega \to -1/4$ all of the graviton KK states completely 
decouple as seen in Fig.~\ref{fig2}. The value of $\Omega$ (and hence 
$\alpha$) can be precisely 
determined at the ILC provided at least the first 2 KK excitations are 
kinematically accessible; this can be done by measurements of ratios of 
the masses and widths of these two states{\cite {me}} which depends only upon 
$\Omega$. It is likely that 
such measurements will be sensitive to $\delta\Omega \sim 0.01$ or better. 
We also see that the KK states now get even more narrow as one moves up the 
KK tower. 

\section{Influence on the ADD Model}

\begin{figure}[htbp]
\includegraphics[width=5.7cm,angle=90]{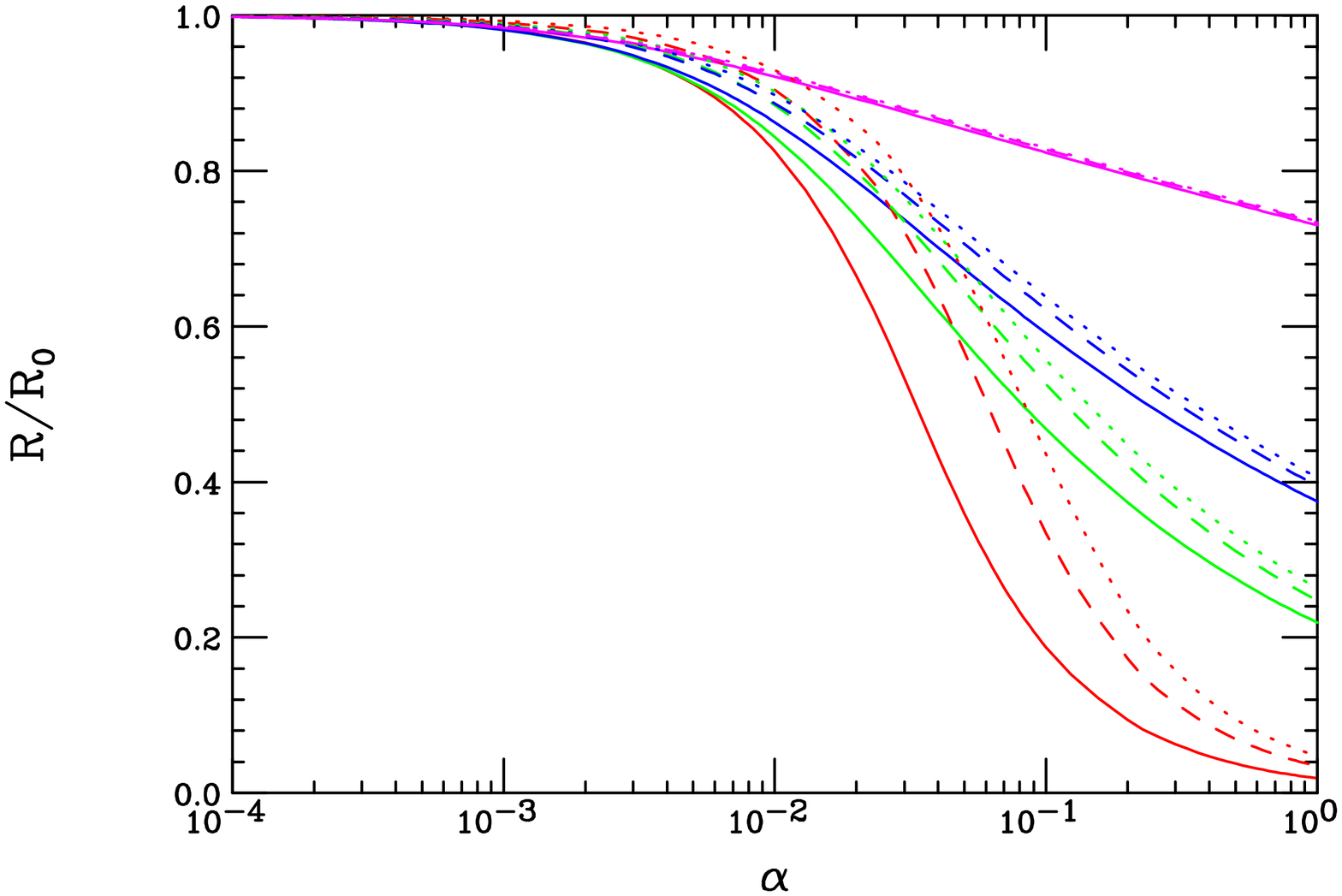}
\includegraphics[width=5.7cm,angle=90]{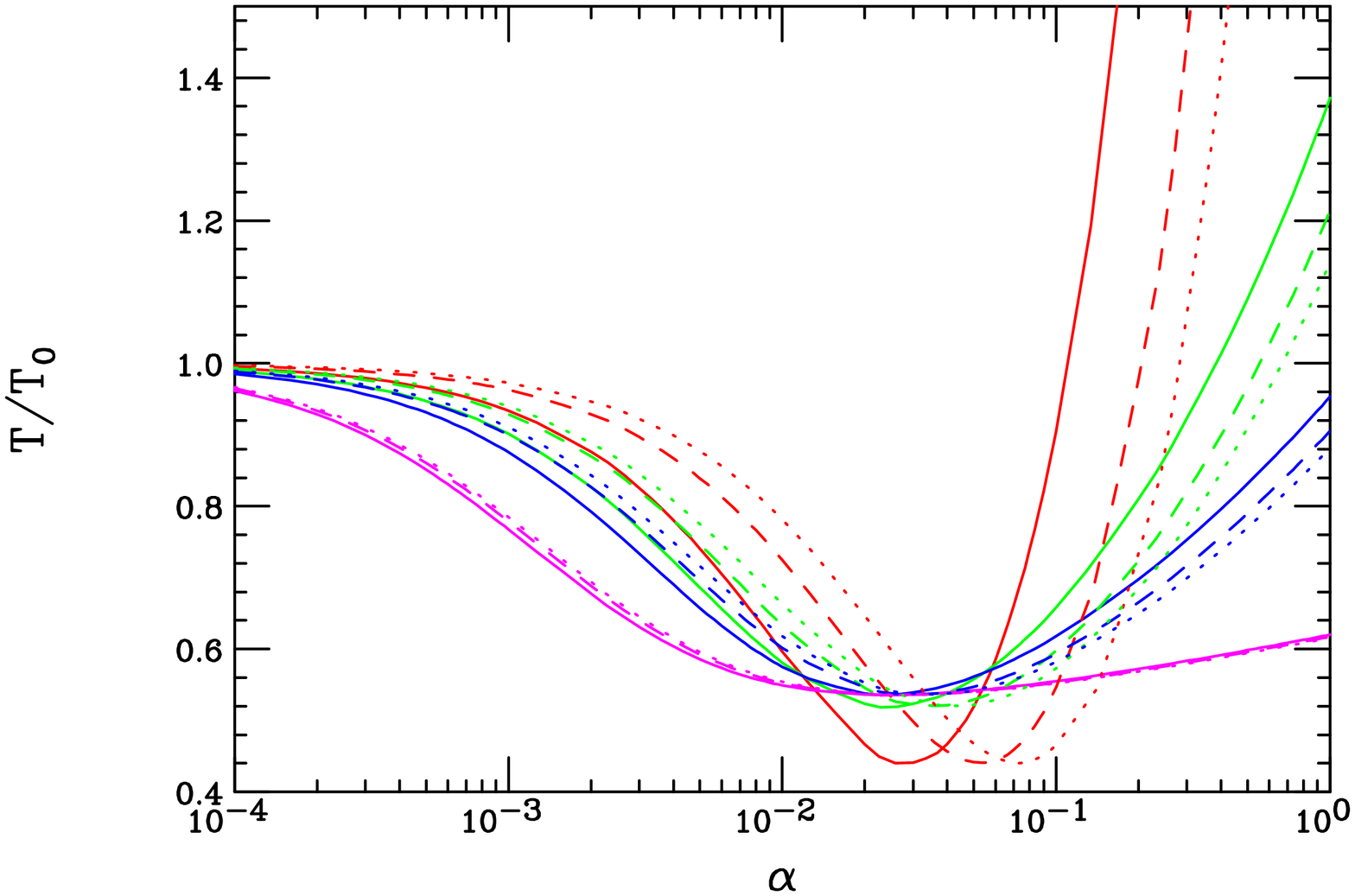}
\vspace*{0.1cm}
\caption{Influence of $\alpha \neq 0$ on the BH mass-Schwarzschild 
radius(left) and temperature(right)  
relationship for $n=2,4,6$ and 20, corresponding to the red, green, blue and 
magenta sets of curves, respectively. The solid, dashed and dotted curves in 
each case correspond to $m=M_{BH}/M_*=2,5,8$, respectively. Here 
$\beta=\gamma=0$ and quantities with an index ``0'' label the predictions 
from the EH action.}
\label{fig3}
\end{figure}

Now let us turn our attention to the ADD case. Since both the bulk and brane 
are flat in this case higher curvature terms do not quantitatively modify 
the usual two signatures of ADD{\cite {JM}}: missing energy from graviton 
emission and new dimension-8 contact interaction operators from KK exchange. 
All of the usual ADD relationships, such as $\mpl^2=V_nM_*^{n+2}$, 
together with the KK graviton mass spectrum and couplings are left completely 
unaltered by the higher curvature terms. The presence of Lovelock terms in 
the action will never be probed by observables associated with such 
processes. The last ADD signature is TeV-scale black hole(BH){\cite {BH}} 
production; 
here we might expect some modifications as the region of 
space near BH are highly 
curved. We remind the reader that BH are expected to form in the collision 
of two partons once energies above $\sim M_*$ are reached with a cross section 
given by 
$\hat \sigma \simeq \pi R^2 \theta(\sqrt {\hat s}-M_*)$,  
where here $R$ is 
the D-dimensional Schwartzschild radius corresponding to the value of 
$M_{BH}\simeq \sqrt {\hat s}$; note the unphysical step function turn-on. 
This cross section 
does not correctly model the turn on of BH production but assumes that it 
starts immediately once $M_*$ is reached. Once formed,  
the BH, now described by a temperature $T$, should decay rapidly by Hawking 
radiation into a number of SM particles. 
The presence of Lovelock terms in the action alters the usual relationships
between the BH mass, radius and temperature, \eg, allowing for 
Lovelock terms, $R$ is obtained by solving 
\begin{eqnarray}
M_{BH}/M_*&=&c\Big[x^{n+1}+\alpha n(n+1)x^{n-1}+\beta n(n+1)(n-1)(n-2)x^{n-3}
\nonumber \\
 &+&\gamma n(n+1)(n-1)(n-2)(n-3)(n-4)x^{n-5}\Big]\,,
\end{eqnarray}
where $x=M_*R$ and $c$ is a constant given by 
$c={{(n+2) \pi^{(n+3)/2}}\over {\Gamma({{n+3}\over {2}})}}$. Other BH 
properties are also modified by the Lovelock parameters as we will discuss 
below. The 
influence of a non-zero $\alpha$ on $R$ and $T$ are explicitly 
shown in Fig.~\ref{fig3}. 
Here we see that O(1) corrections to the usual BH quantities are possible  
for non-zero $\alpha$. 
{\it If} we think of the Lovelock terms as arising from a perturbative-like 
expansion of the full 
action, as in string theory, the coefficients must grow smaller for the higher 
order terms. When we examine the expressions above we see 
that for a perturbative expansion to make sense we must have all of 
$\alpha n^2,\beta n^4$ and $\gamma n^6<1$. Since $n$ can 
be as large as 6 within the usual ADD scenario  
we might expect that $\alpha \sim 10^{-2}$, 
$\beta \sim 10^{-3}-10^{-4}$ and $\gamma \sim 10^{-5}$ with  
wide margins allowed for errors in these estimates. In the ADD case we will 
assume that the Lovelock parameters are positive quantities. 

\begin{figure}[htbp]
\includegraphics[width=5.7cm,angle=90]{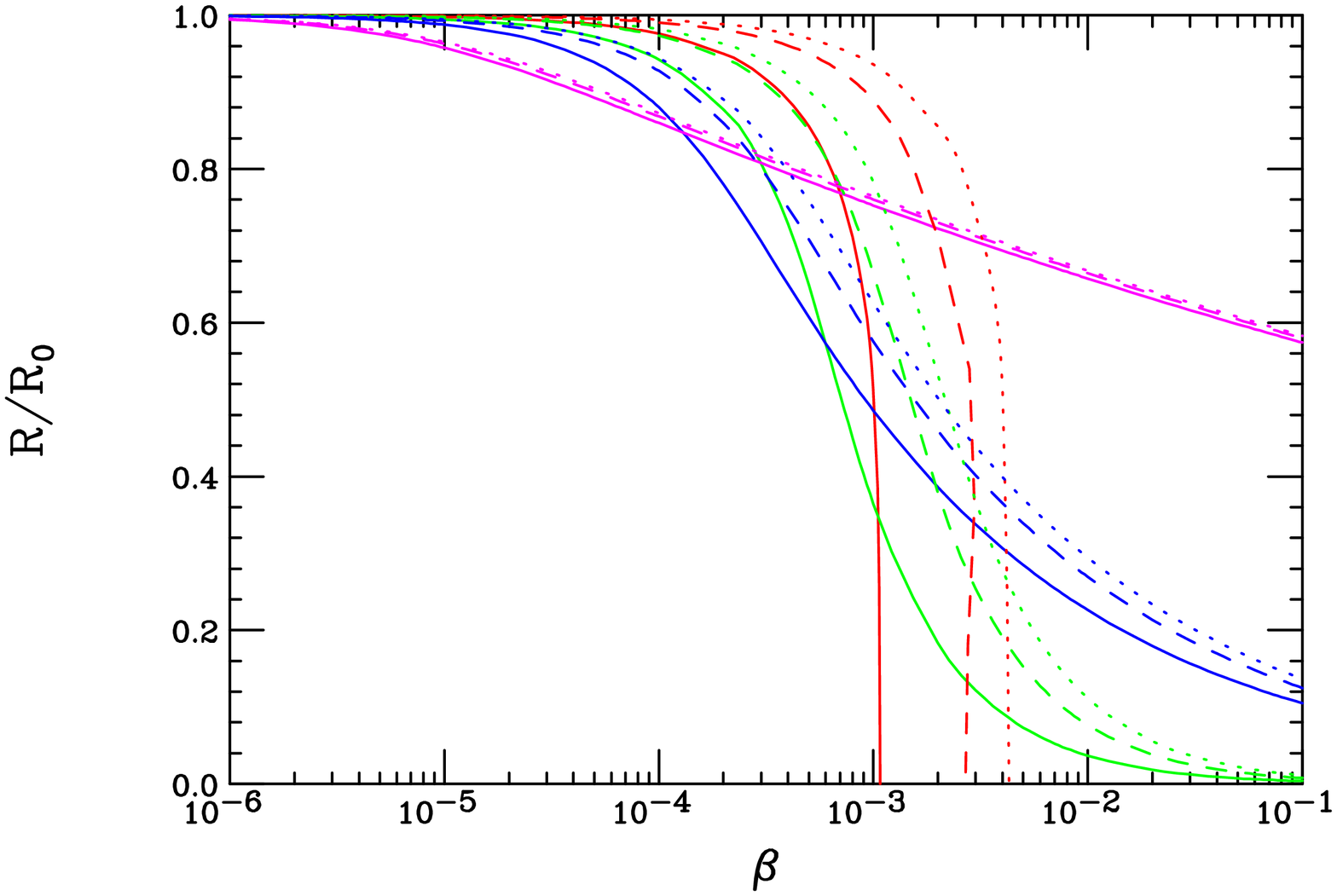}
\includegraphics[width=5.7cm,angle=90]{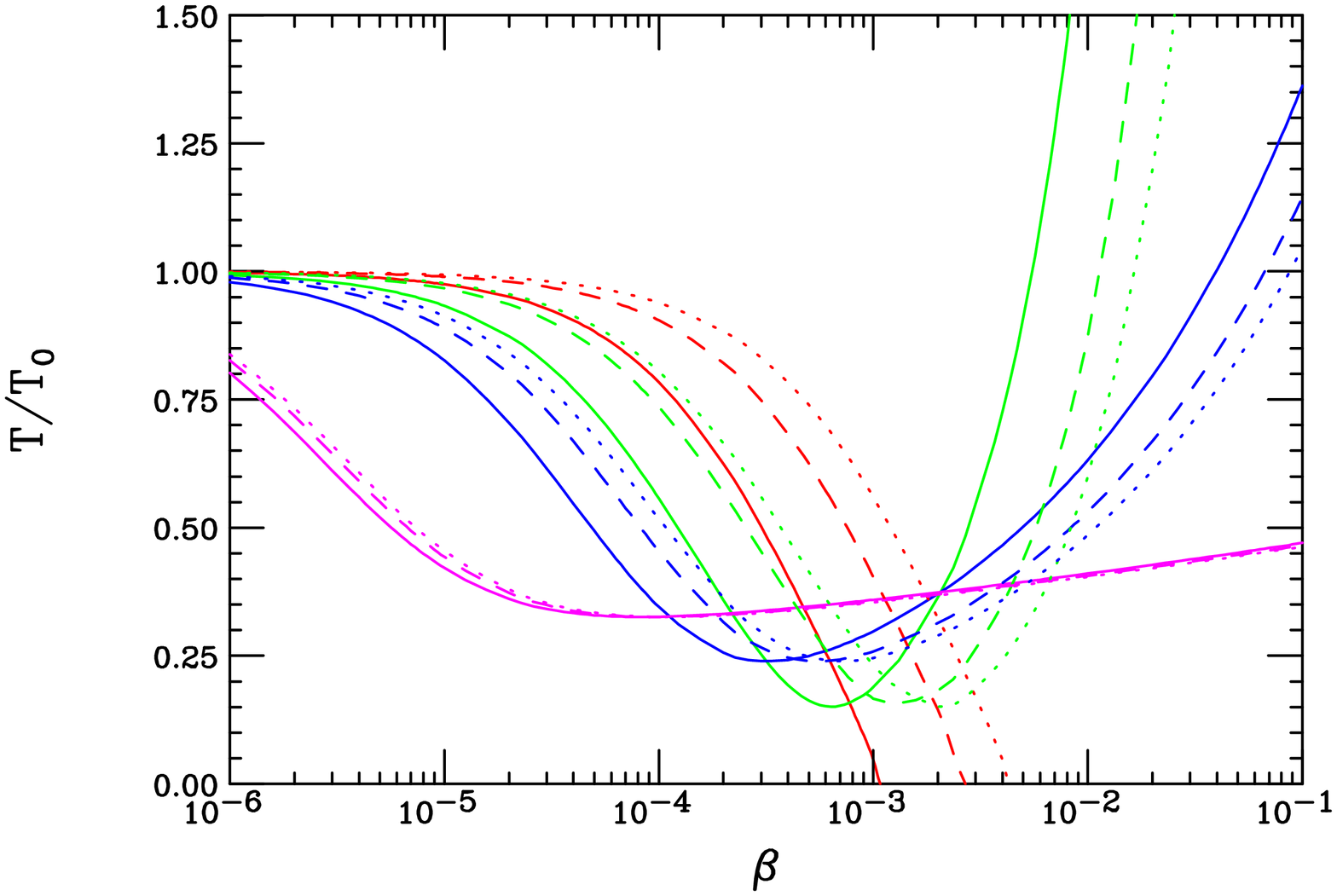}
\vspace*{0.1cm}
\caption{Same as the previous figure but now as a 
function of $\beta$ with $\alpha=\gamma=0$ and $n=2 \to n=3$ since 
${\cal L}_3$ vanishes when $n=2$.}
\label{fig4}
\end{figure}

When $\beta(\gamma)$ becomes non-zero, new {\it qualitative} changes in 
BH properties become possible for the case of $n=3(5)$ as is shown in 
Fig.~\ref{fig4}. Here we see that with $n=3$ for a critical value of $\beta$, 
both $R$ and $T$ are driven to zero. In fact one can show that a threshold 
behavior occurs, \ie, for $n=3$, unless $M_{BH}>60\pi^3\beta M_*$ no BH 
horizon forms; for $n=5$ the threshold occurs at $840\pi^4\gamma M_*$. 
Furthermore, for  
values of $\beta$ and $\gamma$ in the `natural' ranges discussed above this 
tells us that BH will not form below a critical center of mass energy at a 
collider. The value of this mass threshold as well as the shape of the BH 
production cross section immediately above threshold are determined solely 
by the Lovelock parameters. 

\begin{figure}[htbp]
\includegraphics[width=5.7cm,angle=90]{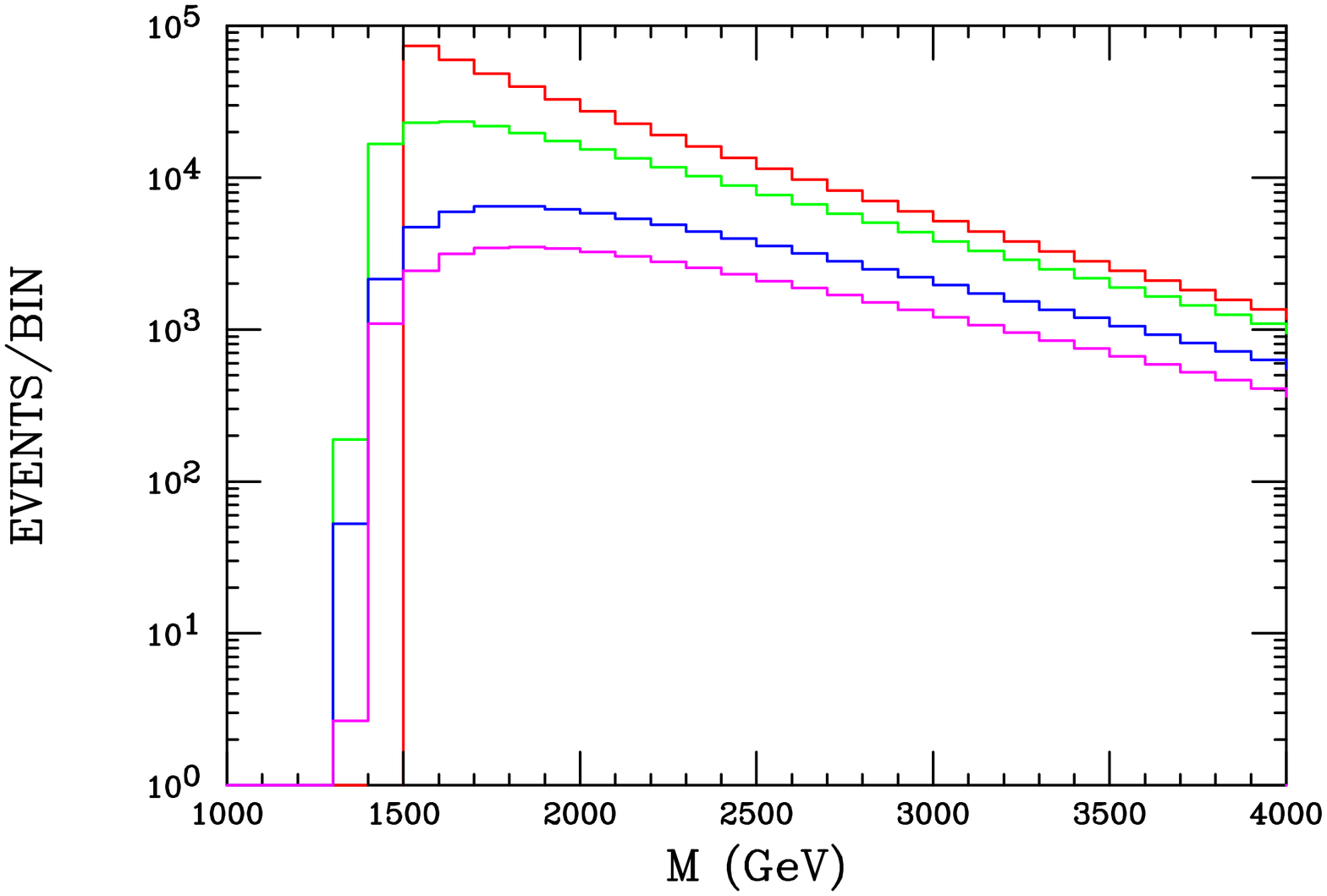}
\includegraphics[width=5.7cm,angle=90]{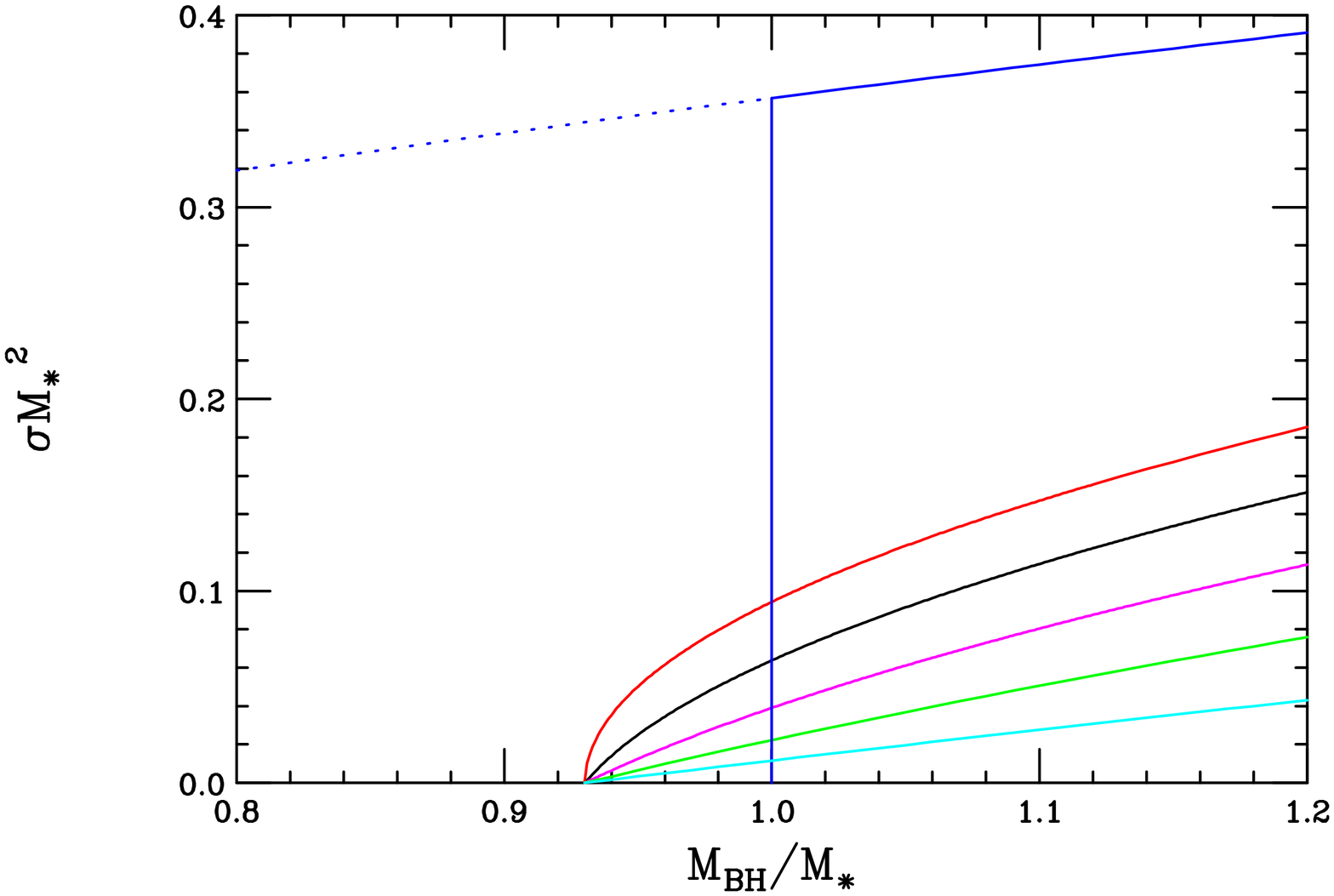}
\vspace*{0.1cm}
\caption{Threshold behavior of the BH cross section at the LHC(left) and 
ILC(right) for $n=3$ and $\beta=0.0005$. In the LHC case the top curve 
assumes the absence of Lovelock terms, while the subsequent ones correspond 
to the above $\beta$ value with 
$\alpha=0(0.01,0.02)$; a luminosity(bin width) of 100 
fb$^{-1}$ (100 GeV) has been assumed. In the ILC case a scaled cross section 
is presented with $\sqrt s=M_{BH}$. Here, 
the top blue line in the naive $\theta$-function that is 
usually assumed in the absence of Lovelock terms. The subsequent   
curves correspond to $\beta$ as given above with 
$\alpha=0,0.002,0.005,0.01$ and 0.02 from top to bottom, 
respectively}
\label{fig5}
\end{figure}

These threshold shapes can be measured at the LHC as shown in 
Fig.~\ref{fig5} but precision measurements will require the ILC. It is 
important to notice that in both cases the cross sections are quite large 
$\sim 10-100$ pb which may yield up to $10^5-10^6$ events so that there is no 
shortage of available statistics. We note that asymptotically, far above 
threshold, the cross 
section we obtain becomes those of the simple step function model. 
Examining Fig.~\ref{fig5} we see that the ILC 
should be able to precisely determine the various Lovelock parameters with 
high precision. We note that these thresholds for BH production do not form 
when $n$ is even.  

\begin{figure}[htbp]
\includegraphics[width=5.7cm,angle=90]{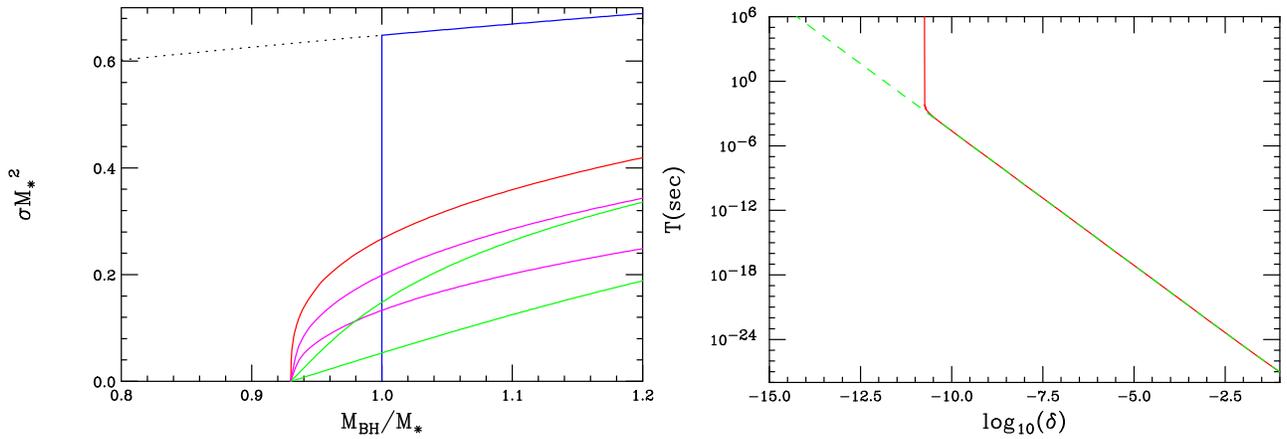}
\includegraphics[width=5.7cm,angle=90]{time.ps2}
\vspace*{0.1cm}
\caption{(Left)Close up of the BH production threshold at the ILC for 
$n=5$; all curves have $\gamma \simeq 1.14\cdot 10^{-5}$ and, from top to 
bottom, have $\alpha,\beta=0,0(0.003,0; 0,0.00003; 0.01,0; 0,0.0001)$, 
respectively. (Right) Decay of an $n=3$ BH with $\beta=0.0005$ and an initial 
mass of 1.5 times the threshold value(green dash). Here $M_{BH}/M_*=
M_{thresh}/M_*+\delta$.}
\label{fig6}
\end{figure}

The BH with $n=3,5$ that we have been discussing have another interesting 
property. BH that result from the EH action have a negative heat capacity. 
As the EH BH evaporate by Hawking radiation and lose mass they become hotter 
and evaporate more quickly until they are completely gone. This process 
occurs quite rapidly for TeV-scale BH, $\sim 10^{-25}$ sec or less. In the 
Lovelock case for $n=3,5$, the heat capacity is positive so that BH will cool 
as they lose mass. In this case they will Hawking radiate until they 
reach the threshold mass where they become (semi-classically!) stable, \ie, 
if we start with a BH with a mass of, say, 1.5 times the threshold value and ask 
how long it will take for it to Hawking radiate down to the threshold mass we 
obtain infinity. This can be seen in Fig.~\ref{fig6}. It would be interesting 
to study the properties of these long-lived BH if they were embedded in media 
of various densities. 

\section{Summary and Conclusion}

In this note we have briefly 
summarized the influence of higher-order Lovelock curvature 
terms on the phenomenology of the familiar RS and ADD models. In the case of 
RS model the dominant effect is a modification to the Kaluza-Klein graviton mass 
spectrum and their associated couplings to matter on the TeV brane. The mass 
spacing between the KK states increases, their couplings become KK level 
dependent and weaker in overall strength. Since KK masses and widths can be 
measured with very high precision at the ILC given sufficient center of mass 
energy, the value of the single possibly non-zero Lovelock parameter in this 
case should be well determined. 

In the case of the ADD model the modifications are quite different. First, the 
usual ADD signatures, \ie, missing energy and dimension-8 contact interactions, 
remain unaltered at the quantitative level. The presence of Lovelock invariants 
in the action does modify the production as well as the properties of TeV scale 
black holes that are produced with large cross sections in high 
energy collisions 
in this scenario. Similarly to the shifts in the KK properties in the RS model 
there are comparable O(1) modifications in the nature of BH due to Lovelock 
terms. There are, however, some qualitative changes for the case of $n$ odd: 
long-lived BH are possible and a mass threshold now exists 
below which BH horizons will not form. Both of these possibilities 
can be probed in detail at future colliders and it may be possible to determine 
the values of the Lovelock parameters with reasonable precision by measuring 
the shape of the BH production cross section near threshold.

\begin{acknowledgments}

The author would like to thank JoAnne Hewett and Ben Lillie for discussions 
related to this study. 
Work supported by Department of Energy contract DE-AC02-76SF00515.
\end{acknowledgments}

\end{document}